# Zipf-Gramming: Scaling Byte N-Grams Up to Production Sized Malware Corpora

Edward Raff
edward.raff@crowdstrike.com
CrowdStrike
Austin, USA

Ryan R. Curtin
Derek Everett
Robert J. Joyce
curtin_ryan@bah.com
everett_derek@bah.com
joyce_robert2@bah.com
Booz Allen Hamilton
McLean, USA

James Holt
Laboratory for Physical Sciences
College Park, USA
holt@lps.umd.edu

## Abstract

A classifier using byte n-grams as features is the only approach we have found fast enough to meet requirements in size (sub 2 MB), speed (multiple GB/s), and latency (sub 10 ms) for deployment in numerous malware detection scenarios. However, we've consistently found that 6-8 grams achieve the best accuracy on our production deployments but have been unable to deploy regularly updated models due to the high cost of finding the top-k most frequent n-grams over terabytes of executable programs. Because the Zipfian distribution well models the distribution of n-grams, we exploit its properties to develop a new top-k n-gram extractor that is up to 35× faster than the previous best alternative. Using our new Zipf-Gramming algorithm, we are able to scale up our production training set and obtain up to 30% improvement in AUC at detecting new malware. We show theoretically and empirically that our approach will select the top-k items with little error and the interplay between theory and engineering required to achieve these results.

## 1 Introduction

Malware detection and analysis involves numerous stages and pipelines due to the wide range of costs and risks involved. Our team developed and deployed a byte n-gram-based model for malware detection that has been used for several years within the U.S. Federal government [15]. The high-level approach is simple: extract n-grams from a large corpus of malware and benign programs, select the top-k most frequent grams as an initial feature set, and then use an $L_1$ or Elastic-Net penalized model to down-select the features and produce the classifier. This model's simplicity is the source of its value in production, coming with two specific mission-critical benefits: **(1)** The model has no required parser dependencies and cannot "fail" in parsing any individual file. **(2)** The feature extraction and classification of files is so small and fast that it can be done "at line speed", where the bottleneck is in getting the data into a computer system, rather than in computational effort.

The first benefit is needed for reliability in production; the tool can be safely deployed on a variety of systems that are of varying ages, software platforms, and language support. The inference code can be written in a few lines of C, Python, Java, etc., and can thus support a wide breadth of operational environments that occur within the U.S. government. The second benefit is critical in the application, as the volume of data that requires analysis is too great for any kind of parser; even simple disassembly parsers are too expensive, let alone more complex types of "lifting" like decompilation. Because byte n-grams can be extracted and evaluated against a linear model quickly, the model can be used as a first-pass filter to identify a smaller subset of files that need further analysis. This could be in the form of a more complex model, algorithm, feature extractor, or professional analyst for review. A wide array of needs and scenarios exist [35, 19, 10, 22, 52, 55, 27, 18], from simple "prevent infections" to forensic investigations to "find any possible malware", creating a range of FPR/TPR thresholds, for which we recommend each downstream user to adjust the threshold of the decision to their operational needs.

Our issue is in the process of extracting the top-k most frequent n-grams itself, which has become a computational burden preventing timely and effective model updates – a requirement for malware detection systems as malware evolves as an adversarial process. To resolve this, we develop an improved algorithm for extracting the top-k n-grams under the assumption they come from a power law distribution, specifically the Zipfian distribution, to safely and stochastically avoid work while arriving at a solution of equal quality.

## 2 Related Work

Machine learning for malware is challenging for many reasons, including concept drift and the need to update models frequently [61, 42]. This has been a challenge for us to update this model as we were unable to increase the size of the training set and still provide timely updates. In our production model, we use multiple evaluation splits based on time to mitigate this and better determine the utility in deployment [37] so that we can have a more accurate estimation of real-world performance. Our use of byte-based features has been widely studied as the exclusive feature representation, but always on smaller datasets [45, 67, 12, 3, 16, 60, 65, 14, 17], and byte-based features have still been valuable to production models that perform parsing or feature extraction [56] or improve Transformer-based approaches in industry [51]. Popular perception is that AntiVirus (AV) systems do not use such simple methods due to advertisement of their ML capabilities, but reverse engineering shows that they still depend extensively on simpler approaches like signatures [4], of which byte n-grams are a major subset [46, 24, 49, 13]. The goal of this article is not to design a complete Anti-Virus system but to improve our faster model used to pre-filter files as the only viable option to getting some kind of result on all production data [34, 33].

As we look at byte n-grams, it is notable that power laws naturally arise in text [25, 38, 53] citation rates [48] code repositories [26], where there are a few items with high frequency and many items with exceptionally low frequency. Zipf's law or the Zipfian distribution is frequently relevant when the number of items is finite [32]. This is useful empirically and theoretically, as it eliminates the challenge of fitting the most correct power law given that other popular options are either continuous or for infinite support [7], which means we can, by the process of elimination, focus only on the Zipf distribution. By focusing on the worst-case Zipf, i.e., the parameterization with the weakest power law, we eliminate the need to estimate any parameters to obtain useful theory. Although the power-law nature of n-grams has been recognized, it has not been leveraged in feature extraction code. Relevant works in n-gramming have found it still a competitive baseline with deep-learning when properly tuned (that is much faster, a key need in our case) [57]. N-grams are also still critical in certain domains such as bioinformatics [68], where our 8-grams over a byte (256 value) alphabet could be adapted to support 32-grams over the 4 value alphabet in genome sequencing (or 64-grams if we used 128-bit long longs). Notably, finding the top-k most frequent n-grams (also called "k-mers" in bioinformatics) is itself a common need for n-grams in the range [20, 60] [54]. While still important to bioinformatics, most popular tools perform chunking or caching to reduce in-memory use, but still naïvely count n-grams and filter after [50]. Though our approach is relevant to the task, k-mers also have a "quality score" for each item in the sequence that is beyond the scope of our work, where bytes are known and exact.

## 3 Fast Top-k n-Gramming

The models we want to train are byte $n$-grams, so there are $256^n$ possible $n$-grams. We have found that $n \in [6, 8]$ usually performs best, which means there are more possible $n$-grams than bytes in our training data to observe. A model that has any predictive utility must use $n$-grams that re-occur, and so selecting the top-$k$ most frequent $n$-grams is a simple and effective initial step to model construction (i.e., a feature seen only once is improbable in being a helpful feature). Standard ML libraries that support $n$-grams do this in a naïve way: count all observed $n$-grams, and then select the top-$k$ most frequent $n$-grams. This is manageable for $n \in [1, 3]$ as is common in text applications in moderately sized corpora. For our malware datasets that require larger $n$ and are up to 32 terabytes in size, this is a computationally infeasible approach.

Instead, we are concerned with directly estimating the top-$k$ items without holding an unbounded amount of additional memory, because most items are infrequent and observed ≤ 10 times. The canonical algorithm for finding the top-k most frequent items in a stream is the SpaceSaving algorithm [31]. While theoretically beautiful, this data structure achieves low throughput on modern hardware, making it a non-option. It was instead used as a component in the KiloGrams algorithm [40, 43, 41], which performed a two-pass process: **(1)** Estimate the top-k most frequent hashes of n-grams, as the hashes can be enumerated in constant memory. **(2)** On a second pass, use the Space-Saving algorithm only on the white-listed hashes from the first pass, reducing the total number of Space-Saving updates.

The top-k hash strategy is possible only because of the power-law distribution that n-grams tend to follow, as referenced in § 2. Because we have a finite alphabet, the Zipfian distribution (or "Zipf" for short) [69] is assumed for the distribution of n-grams. The top-k hashes will occur with much greater frequency than the lower-ranked/less-frequent items and so make an accurate correspondence to the true top-k items. While an improvement, we find that the KiloGramming algorithm is still too slow due to the large $\approx 2^{31}$ sized table it must create. We will improve upon the KiloGramming algorithm with our Zipf-Gramming approach.

Our Zipf-Gramming approach is shown in Algorithm 1, where blue and green lines correspond to our improvements over the original KiloGramming algorithm. If removed, the remaining black lines constitute the original KiloGramming approach. The key issue with KiloGrams is that updating the large hash table in the first pass (line 10) and the SpaceSaving data structure in the second pass (line 24) is prohibitively slow. Our strategy then is to avoid performing these updates as much as possible and do just enough updates to obtain a reasonable estimate of the true top-k items. In blue, we take advantage of the Zipf distribution to stochastically avoid counting n-grams in a manner that allows us to reach the same top-k items with high fidelity. In green, we create a smaller write/evict cache to act as a buffer to accumulate updates before pushing them to the larger data structures. These algorithmic improvements, combined with careful software engineering to avoid as many even mildly expensive operations as possible, combine to produce the first top-k n-gram algorithm that can process large datasets at gigabit speeds and greater for large datasets.

In the remainder of this section, we will first detail more explicitly and empirically why the KiloGramming approach is slow and how the hash-table results in worst-case hardware characteristics in § 3.1. Then, we will theoretically justify our stochastic counting approach in § 3.2. Because the write/evict cache is sensitive to implementation details, its proofs and discussion will occur in § 4.

**Algorithm 1** Zipf-Gramming, blue and green are additions from the original Kilo-Gram algorithm

---

**Require:** Bucket size $B$, rolling hash function $h(\cdot)$, corpus of $C$ documents, and desired number of frequent items $k$.
1: $B \leftarrow$ new integer array of size $|B|$
2: Create Cache $C$ of size $|C| = k$
3: Determine $p_{\text{skip}}$, the probability of skipping the given token.
4: **for** all documents $x \in C$ **do**         ▷ $O(L)$ for $L$ total $n$-grams
5:   **for** $n$-gram $g \in x$ **do**
6:     **if** $p_{\text{skip}} < p \sim \mathcal{U}(0, 1)$ **then**
7:       Skip processing this n-gram.
8:     $q' \leftarrow h(g) \mod B$
9:     **if** $q' \neq \text{Key}(C[q'])$ **then**   ▷ Check if g is in the cache
10:       $B[q'] \leftarrow B[q'] + \text{Value}(C[q'])$
11:       $\text{Key}(C[q']) \leftarrow q'$
12:       $\text{Value}(C[q']) \leftarrow 0$
13:     $\text{Value}(C[q']) \leftarrow \text{Value}(C[q']) + 1$
14: $B_k \leftarrow \text{QuickSelect}(B, k)$
15: $S \leftarrow$ new Space Saving structure with $B_S$ buckets.
16: Refine $p_{\text{skip}}$, the probability of skipping the given token.
17: **for** all documents $x \in C$ **do**         ▷ Second pass over data
18:   **for** $n$-gram $g \in x$ **do**
19:     **if** $p_{\text{skip}} < p \sim \mathcal{U}(0, 1)$ **then**
20:       Skip processing this n-gram.
21:     $q' \leftarrow h(g) \mod B$
22:     **if** $q' \in B_k$ **then**
23:       **if** $q' \neq \text{Key}(C[q'])$ **then**   ▷ Check if g is in the cache
24:         Insert $\text{Key}(C[q'])$ into $S$ $\text{Value}(C[q'])$ times
25:         $\text{Key}(C[q']) \leftarrow g$
26:         $\text{Value}(C[q']) \leftarrow 0$
27:       $\text{Value}(C[q']) \leftarrow \text{Value}(C[q']) + 1$
28: **return** top-$k$ entries from S

### 3.1 Bottlenecks of the KiloGramming

Although the *KiloGramming* algorithm produces a significant speedup compared to the naïve approach of maintaining a list of $n$-grams with counts, it suffers from some significant bottlenecks when implemented on real-world hardware. Primarily, the bottleneck is associated with $B$, the array used in the first pass that holds the counts for each hash value. Because the bucket size $B$ is typically chosen to be a large value, such as $2^{31} - 19$, this means the amount of RAM required to hold $B$ is 8 GB if 32-bit integers are used, and 16 GB if 64-bit integers are used. For modern systems, this is not an onerous requirement: RAM sizes are often 64 GB and larger.

But the memory access patterns of the *KiloGramming* algorithm are highly problematic: because each $n$-gram is hashed to a bin in $B$ uniformly randomly, each successive update to $B$ requires accessing a random location in $B$. This is a worst-case situation for memory access, as the hardware prefetcher cannot predict the next access location. As a result, the memory access request can only be issued once the hash value $h(g)$ is known, causing significant latency for each individual update.

Worse yet, because $B$ is so large, it spans many pages of virtual memory. Before a memory access request can succeed, the virtual page must be mapped to physical memory. Virtual page mappings are cached by the translation lookaside buffer (TLB). If a memory access request corresponds to a virtual page that is *not* cached in the TLB—which is going to be virtually every lookup in $B$ because we are accessing $B$ fully randomly—a walk of the page table must occur. Each of these walks can be very expensive.

It is possible to use a small circular buffer to mask this latency somewhat by manually issuing a software prefetch command for $B[h(g)]$ and storing $h(g)$ in the circular buffer, and then actually performing the increment of $B[h(g)]$ on the next pass through the circular buffer. But on many processors, including recent Intel processors [8], prefetch commands will stall on a TLB miss, synchronously performing a page table walk. Only once the page table walk is completed will the prefetch be issued.

So, overall, the memory access pattern required by the *KiloGramming* algorithm is basically worst-case: random accesses in memory mean that hardware prefetchers are ineffective; the huge size of $B$ means that many pages of memory are required, meaning constant TLB misses and page walks. This causes *KiloGramming*'s main bottleneck to be updates to $B$ instead of reading the underlying data from disk. It is easy to demonstrate the significant effect that these memory access patterns have. Table 1 presents a simple example inspired by Drepper [9]: we generate an array of a few different sizes, and increment elements sequentially as well as randomly. The *KiloGramming* situation—random access, largest array size—runs at $\sim 100$ **MB/s**! This is many orders of magnitude away from peak memory bandwidth and is even a couple orders of magnitude away from typical modern disk speeds.

| Size of array | Sequential | Random |
|---|---|---|
| 4 kB | 27.1 GB/s | 0.56 GB/s |
| 2 MB | 32.5 GB/s | 0.54 GB/s |
| 8 GB | 10.0 GB/s | **0.10 GB/s** |

Table 1: At the sizes required for *KiloGramming*, the penalty is orders of magnitude!

Having now identified the bottleneck, we present our simple solution: just access $B$ less often. It turns out that we can do this probabilistically without affecting the quality of the results.

### 3.2 Zipf Counting

We assume our $n$-grams follow a power law distribution, which is expected and previously observed when using n-grams [25, 38, 53, 48, 26]. In particular, we use the Zipf distribution, over an alphabet of $\Sigma$ possible items, with the probability mass function Equation 1. $H_\Sigma^\rho = \sum_{i=1}^{\Sigma} \frac{1}{i^\rho}$ is the harmonic number.

$$\text{PMF}_{\text{Zipf}}(x) = \begin{cases} \frac{x^{-\rho-1}}{H_\Sigma^{(\rho+1)}} & 1 \leq x \leq \Sigma \\ 0 & \text{otherwise} \end{cases} \quad (1)$$

We will use $\rho = 0$ in our analysis as the worst-case possible distribution because it maximizes the weight that is in the tail. See Figure 1 as a demonstration of this.

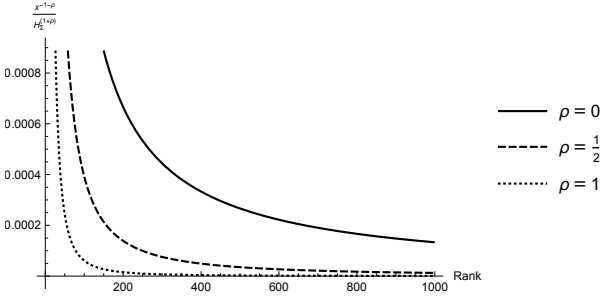

Figure 1: The PMF of the Zipf distribution with $\Sigma = 2^{10}$ as an example for multiple values of $\rho$. The larger $\rho$ is, the less mass in the tail of the distribution, and the easier it is to find the true top-k via sampling. Thus we assume the worst case $\rho = 0$ for our proofs to guard against distribution assumptions.

Rather than process all $L$ n-grams of a sequence, we want to sub-sample a fraction of only $N < L$ items to count, as this would save significant computational effort. Our interest then is in bounding the probability of making an error in selecting the top-$k$ items when we sub-sample the distribution of n-grams.

### 3.2.1 Binomial Treatment of Error.
To start, we wish to know that, given the higher ranked $k_h$, it will be counted more times than a lower ranked $k_l$, where $k_l > k_h$ because the top-ranked item is $k = 1$, and the lowest ranked item is $k = \Sigma$. Given that we will observe $N$ samples from the distribution, we can describe the number of times the $k$'th rank is observed as a binomial distribution $B(N, p) = p^x \binom{N}{x}(1-p)^{N-x}$, where we set $p = \text{PMF}(\text{Zipf}, k)$, giving the distribution of the number of times we expect this to occur[1].

We desire $k_h$ to have a higher observed count than $k_l$, which we can describe as a transformed distribution $Z = b_h - b_l$ where $b_h \sim B(N, \text{PMF}(\text{Zipf}, k_h))$ and $b_l \sim B(N, \text{PMF}(\text{Zipf}, k_h))$. The range of positive values of the distribution, i.e., $\mathbb{P}(Z > 0)$, describes the cases where rank $k_h$ properly has a higher observed count than the lower rank $k_l$. While we can calculate the characteristic function of this distribution as $\left(-\frac{1}{k_h H_\Sigma} + \frac{e^{ix}}{k_h H_\Sigma} + 1\right)^N \left(-\frac{1}{H_\Sigma k_l} + \frac{e^{-ix}}{H_\Sigma k_l} + 1\right)^N$, it does not integrate easily and so we can not directly calculate the value of interest.

A loose bound on the probability of an error can be made by exploiting the linearity of expectation and variance to bound the result using Cantelli's inequality [6] to arrive at Theorem 1. All proofs will be deferred to the appendix.

THEOREM 1. *The probability of a rank $k_l$ item having a count of $\lambda$ more than a higher rank item $k_h$ is* $\leq \frac{N k_h H_\Sigma k_l (k_h + k_l) - N\left(k_h^2 + k_l^2\right)}{k_h^2 \left(H_\Sigma k_l \left(\lambda^2 H_\Sigma k_l + N\right) - N\right) + N k_h H_\Sigma k_l^2 - N k_l^2}$

PROOF.

$$\mathbb{E}[Z] = \mathbb{E}[b_h] - \mathbb{E}[b_l] = \left(\frac{N}{k_h H_\Sigma}\right) - \left(\frac{N}{H_\Sigma k_l}\right) = \frac{N\left(\frac{1}{k_h} - \frac{1}{k_l}\right)}{H_\Sigma}$$

---
[1]Strictly speaking, when considering all $k$, we would need a Multinomial distribution over all options, but the analysis quickly becomes intractable so we accept this small departure for analytic efficacy.

$$\mathbb{V}[Z] = \mathbb{V}[b_h] + \mathbb{V}[b_l] = \left(\frac{N(k_h H_\Sigma - 1)}{k_h^2 (H_\Sigma)^2}\right) + \left(\frac{N(k_l H_\Sigma - 1)}{k_l^2 (H_\Sigma)^2}\right)$$

$$= \frac{N k_h^2 H_\Sigma k_l + N k_h H_\Sigma k_l^2 - N k_h^2 - N k_l^2}{k_h^2 (H_\Sigma)^2 k_l^2}$$

$\mathbb{P}(Z - \mathbb{E}[Z]) \leq \frac{\mathbb{V}[Z]}{\mathbb{V}[Z] + \lambda^2}$ via Cantelli's inequality.

Plugging in $\mathbb{E}[Z]$ and $\mathbb{V}[Z]$ completes the bound of the theorem gives

$$\mathbb{P}\left(Z - \frac{N\left(\frac{1}{k_h} - \frac{1}{k_l}\right)}{H_\Sigma}\right) \leq \frac{\frac{N\left(\frac{H_\Sigma}{k_h} - \frac{1}{k_h^2} + \frac{H_\Sigma k_l - 1}{k_l^2}\right)}{(H_\Sigma)^2}}{\lambda^2 + \frac{N\left(\frac{H_\Sigma}{k_h} - \frac{1}{k_h^2} + \frac{H_\Sigma k_l - 1}{k_l^2}\right)}{(H_\Sigma)^2}}$$

. Simplifying the result then gives the bound of the theorem. □

To illustrate the utility of this result, we calculate the bound's result for a conservative $N = 10^{11}$ observations, with an alphabet size of $\Sigma = 256^8$ for byte 8-grams, with a desired $k_h = 10^6$ for the top 1-million n-grams. The results in Figure 2 show that there is a low probability of being off by $\lambda = 1000$ values, which is a 0.1% error that would be inconsequential in practice. We also note that the bound in Theorem 1 is pessimistic due to the $-\mathbb{E}[Z]$ term, as $\mathbb{E}[Z]$ is positive (it is known that the rank is higher and in expectation will have a higher count). This creates a set of virtual counts/observations that the $b_h$ must probabilistically overcome, and the result can show that it is unlikely for $k_l$ to obtain $\lambda$ more counts for moderately large values of $\lambda$.

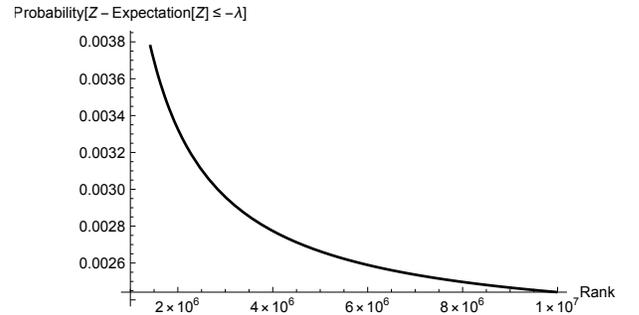

Figure 2: Pessimistic bound on the probability of incorrectly ranking the $k_h = 10^6$'th item below several lower ranks. The result shows that for the parameters of real-world data, we have a low probability of an error for the least frequent but still desired item $k_h$.

This analysis is loose because we can not directly tackle the CDF.

### 3.2.2 Gaussian Approximation Analysis.
The prior result is not ideal because we could not calculate the CDF, which would directly answer the question of $\mathbb{P}(Z \leq 0)$, which corresponds to the probability of $k_l$ getting a higher total count than $k_h$, and thus inappropriately being selected for the top-$k$ n-grams. Tackling the issue by molding it as a binomial distribution is unsatisfying because the value $\mathbb{P}(Z \leq 0)$ which corresponds to the probability of $k_l$ getting a higher total count thank $k_h$ (i.e., probability of an error) is loosely bounded and hard to calculate. Further, the more ideal result is to obtain the CDF for the ordered statistic of all $\min(|\Sigma|, L) - k_h$ distributions for all $k' < k_h$, so that we can quantify the likelihood of *any* lower ranked item receiving a count greater than $b_h$.

We provide positive evidence for both of these questions in this section by leveraging the Gaussian approximation to the binomial distribution, $B(N, p) \approx \mathcal{N}(Np, \sqrt{Np(1-p)})$ which is generally accurate so long as $Np \geq 5$ [36, 47]. In our analysis, this is so that we have a more pliable distribution to analyze, so that we may better understand the reasonableness of our top-k sampling approach (under the caution of using an approximate distribution). Thus, we replace $b_k \sim \mathcal{N}(Np_k, \sqrt{Np(1-p_k)})$ where $p_k = \text{PMF}(\text{Zipf}, k_h)$. This gives a new definition for

$$Z = \mathcal{N}\left(Np_{k_h} - Np_{k_l}, \sqrt{N(1-p_{k_h})p_{k_h} + N(1-p_{k_l})p_{k_l}}\right)$$

., which has the familiar CDF of the Gaussian distribution, which evaluates to:

$$\frac{1}{2}\text{erfc}\left(\frac{Nk_l - k_h(xH_\Sigma k_l + N)}{\sqrt{2Nk_h H_\Sigma k_l (k_h + k_l) - 2N\left(k_h^2 + k_l^2\right)}}\right) \quad (2)$$

We can see from Equation 2 several desired properties. First, the limit as $N \to \infty = 0$, so we can drive the error rate to zero by sampling large values of $N$. (see Figure 3). We can also see the importance of the bounded alphabet size $\Sigma$ due to the limit as $\Sigma \to \infty = 1$.

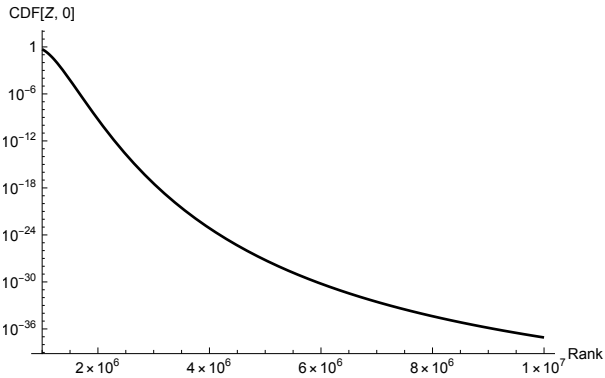

Figure 3: Under the normal approximation to the binomial, we can evaluate the CDF, showing that the probability of an error in ranking drops precipitously as we increase the gap between the item of lowest relevant frequency $k_h$ and the rank of the too infrequent to be selected item $k_l$.

The more challenging question is to bound the likelihood of a range of $k' \geq k_l$ receiving too high a count. Essentially, To bound the likelihood of a range of $k' \geq k_l$ receiving too high a count, let $Z_{k'}^{k_h}$ be defined by the distribution:

$$\mathcal{N}\left(Np_{k_h} - Np_{k'}, \sqrt{N(1-p_{k_h})p_{k_h} + N(1-p_{k'})p_{k'}}\right).$$

What we wish to calculate is $\mathbb{P}(\min\left[Z_{k_l}^{k_h}, Z_{k_l+1}^{k_h}, \ldots, Z_{\min(N,\Sigma)}^{k_h}\right] \leq 0)$, which corresponds to an ordered statistic over the $Z$s. The analytic result can be represented by noting that :

$$\mathbb{P}(\min\left[Z_{k_l}^{k_h}, Z_{k_l+1}^{k_h}, \ldots, Z_{\min(N,\Sigma)}^{k_h}\right] \leq 0) = 1 - \prod_{i=k_l}^{\min(N,\Sigma)} 1 - CDF(Z_i^{k_h}, 0)$$

Substituting in the value for the CDF we obtain

$$\xi_T = 1 - \prod_{k=k_l}^{T}\left(1 - \frac{1}{2}\text{erfc}\left(\frac{\frac{N}{k_h H_\Sigma} - \frac{N}{k H_\Sigma}}{\sqrt{2}\sqrt{\frac{N\left(1-\frac{1}{k_h H_\Sigma}\right)}{k_h H_\Sigma} + \frac{N\left(1-\frac{1}{k H_\Sigma}\right)}{k H_\Sigma}}}\right)\right) \quad (3)$$

Which does not simplify on its own, and ordered statistics of non-identical distributions are particularly challenging [2]. Our approach is to upper-bound this distribution by replacing all $Z_{k_l+c}^{k_h}$ with $Z_{k_l}^{k_h}$, which is pessimistic but useful in judging a range of gaps, yielding Theorem 2.

THEOREM 2. *The expected difference in observations between $h_k$ and the $T$ items after the $h_l$'th lower ranked item is*

$$\geq -\sqrt{2}\sqrt{\log(T)}\sqrt{\frac{N\left(1-\frac{1}{h_k H_\Sigma}\right)}{h_k H_\Sigma} + \frac{N\left(1-\frac{1}{h_l H_\Sigma}\right)}{h_l H_\Sigma}} + \frac{N}{h_k H_\Sigma} - \frac{N}{h_l H_\Sigma}$$

PROOF. Given $T$ i.i.d. random variables $X_i \sim \mathcal{N}(0, \sigma^2)$, we wish to calculate $\mathbb{E}[\min X_1, \ldots, X_T] = -\mathbb{E}[\max X_1, \ldots, X_T]$ due to symmetry. Taking $\mathcal{X}_{\max} = \mathbb{E}[\max X_1, \ldots X_T]$ one can use Jensen's inequality to obtain $\mathbb{E}[\mathcal{X}_{\max}] \leq \frac{\log T}{t} + \frac{t\sigma^2}{2}$. The minimizer of $t = \frac{\sqrt{2\log n}}{\sigma}$ yields the result that $\mathbb{E}[\mathcal{X}_{\max}] \leq \sigma\sqrt{2\log T}$. Applying the sign change for the minimum produces $\mathbb{E}[\mathcal{X}_{\min}] = \mathbb{E}[\min X_1, \ldots, X_T] = -\mathbb{E}[\mathcal{X}_{\max}] \geq -\sigma\sqrt{2\log T}$. Next, we note that our interest is in

$$\mathbb{E}\left[\min\left[Z_{k_l}^{k_h}, Z_{k_l+1}^{k_h}, \ldots, Z_T^{k_h}\right]\right] \leq \min\left[\overbrace{Z_{k_l}^{k_h}, Z_{k_l}^{k_h}, \ldots, Z_{k_l}^{k_h}}^{T \text{ times}}\right], \text{the later}$$

bound can be replaced by $\mathbb{E}[\mathcal{X}_{\min}] + \mathbb{E}[Z_{k_l}^{k_h}]$ due to the linearity of expectations, giving $\mathbb{E}[\min\left[Z_{k_l}^{k_h}, Z_{k_l+1}^{k_h}, \ldots, Z_T^{k_h}\right]] \leq -\mathbb{E}[\mathcal{X}_{\max}] \geq -\sqrt{\mathbb{V}[Z_{k_l}^{k_h}]}\sqrt{2\log T} + \mathbb{E}[Z_{k_l}^{k_h}]$. Substituting the previously derived variance and expectation of the Gaussian approximation of the binomial of interest results in the theorem's bound. □

We use Theorem 2 to demonstrate that even over an order of magnitude of lower-ranked items, we still expect, on average, the approach of sub-sampling Zipfian distribution to accurately select from the top-k items. In Figure 4, we plot this bound normalized by the least frequent but desired item $k_h$'s expected count (i.e., how much excess buffer in counts occurs). We see that this value is positive, and the buffer increases as we choose larger thresholds $k_h$.

This theorem also allows us to explore the trade-offs in accuracy. This is simulated in Figure 5, as we increase the tail of the distribution and the sample size $N$. The results show that the sample size $N$ dominates the error rate, indicating that if insufficient performance occurs in practice, we should increase the sample rate $N$, and that the number of items itself should be a di minimus factor provided the data follows a Zipfian distribution.

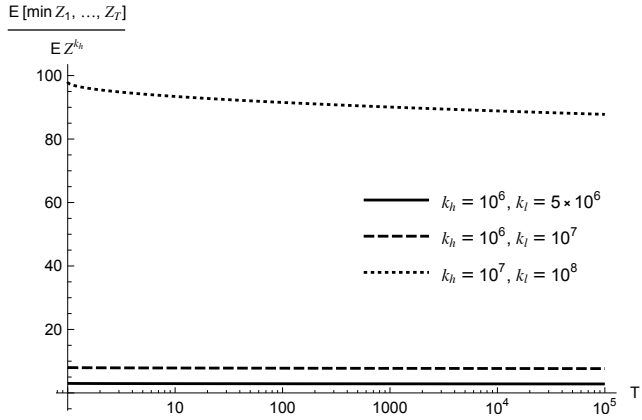

Figure 4: The y-axis shows the ratio of excess expected counts to select the $k_h$'th item over the preceding cumulative $T$ items (x-axis). Using relevant real-world values of $\Sigma = 256^6$ and $n = 10^{12}$, we see that we always expect a positive value (and thus no errors on *average*).

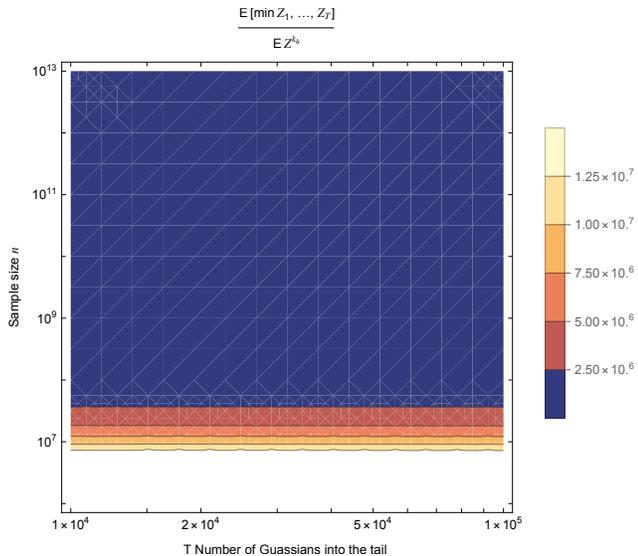

Figure 5: Using our bound, we look at how both the sample size (y-axis) and number of lower ranks (x-axis) impact the results (color, note all values are positive meaning the expected result is always the desired positive ranking of $k_h$ above $k_l$, for various offsets $k_l = k_h + T$). The dominant factor is the number of samples $N$ taken.

## 4 Efficient Implementation

As a motivating example of the need to be careful in implementation, consider that a modern high-end 5 GHz CPU can perform $\approx 43{,}000$ million instructions per second, which is (roughly) 43 instructions per nanosecond[2]. A top-of-the-line NVMe SSD can obtain a theoretical maximum throughput of $\approx 7$ GB/s, or 7 bytes per nanosecond. While both numbers are theoretical maximums that cannot usually be achieved, it puts into context that at optimal compute efficiency, just moving bytes constitutes 16% of all available compute to obtain maximum data transfer rates. As we illustrated in § 3.1, the compute portion of the original KiloGram process is exhibiting the worst-case behavior. Thus, any complex data structure update (like SpaceSaving) is prohibitively expensive and needs to be minimized, and any non-update states must be highly efficient to make up for the lost time on updates. This is especially important in the first pass of the algorithm when there is no whitelist to further reduce the data structure update rate.

We rely heavily on bit-twiddling operations like AND, OR, and XOR that have a latency of 1 cycle in modern hardware, and integer multiplications are usually only 3-5 cycles. But even a modulo operation (e.g., `hash % range`) to index into a table can be too expensive at 15-71 cycles. Floating point operations can vary from 9-400 cycles, and so must be avoided at all costs [11].

In this section, we will review low-level details and implementation decisions that we found were required to reach high data throughput. This includes faking a "rolling" hash function given that we are skipping most bytes, how to avoid floating point math for determining the next byte to evaluate, and the choice behind and proof of our write/evict cache strategy.

### 4.1 Rolling Hash Function $h(\cdot)$

The original KiloGram's implementation used the Rabin Karp rolling hash function (RKH). Given a sequence of $a_1, \ldots, a_n$ tokens $h(\cdot) = a_1 + a_2 B + \ldots + a_n B^n$ where $B$ is a (generally prime) constant. When one wants to shift or "roll" the has function over by one token to $a_{n+1}$, you can calculate this in constant time as $B \cdot h(\cdot) + a_{n+1} - B^n \cdot a_1$. By pre-calculating $B^n$, this can be updated using only two integer multiplications and two integer additions.

However, we found that this hash function is prohibitively slow in practice due to low computational throughput. This has been observed by others in the HPC and software engineering space, where the naïve RKH can only run at 0.75 GB/s, absent any other operations. This is far below the 7 GB/s theoretical IO throughput of our hardware and optimistic as it excludes any work done around/with the hash values themselves [23, 58]. For example, even though we will skip some number of bytes with probability $p_{\text{skip}}$, the RKH hash requires us to still calculate the hash value for skipped tokens because the hash for the $t$'th token in a sequence is dependent on the $n - 1$ previous tokens.

Instead, we use a 64-bit integer to keep a running estimate of the current "key" n-gram. This can be maintained by a simple bit shift and logical and operation at each new input token; $key = key \ll l | a_{n+1}$ is sufficient to calculate this with highly efficient bitwise operations. Only when we are at a location selected to be counted do we calculate a hash value from the current $n$-gram $key$. For the hash function, we use the SplitMix hash function [59], which has good uniformity of distribution, and requires only 6 bit-twiddles and two integer multiplications. Thus, it is of comparable cost to RKH, but only when we need a hash with better statistical distribution.

```
uint64_t SplitMixHash(uint64_t x) {
```

---
[2]e.g., see https://hothardware.com/reviews/intel-core-i9-11900k-core-i5-11600k-rocket-lake-s-review

```
    x = (x ^ (x >> 30)) * UINT64_C(0xbf58476d1ce4e5b9);
    x = (x ^ (x >> 27)) * UINT64_C(0x94d049bb133111eb);
    x = x ^ (x >> 31);
    return x;
}
```

## 4.2 Skipping Tokens & Geometric Distribution

In lines 6-7 and 19-20 of Algorithm 1, we select with probability $p_{\text{skip}}$ to skip the current item for processing. This optimization is critical for avoiding cache misses when indexing into the large matrix on line 10, and the high computational cost of the Space-Saving algorithm on line 24. However, running any Pseudo-Random Number Generator (PRNG) on every byte sequence is a non-trivial computational cost of its own.

In an ideal world, we would know how many bytes $s$ to skip forward to the next $n$-gram, which would allow us to skip to the next $s - n + 1$ bytes so that we can ensure the *key* storing the current gram $g$ is correctly populated (i.e., you must record the $n - 1$ tokens to know the current $n$-gram). Naïvely, we could set $s = \lfloor 1/p_{\text{skip}} \rfloor$, but this would result in a biased distribution of byte values. This is especially problematic for malware, where content is often aligned to various power-of-two multiples due to a variety of technical reasons.

THEOREM 3. *For an arbitrary threshold $c \in (0, 1)$, the inverse CDF of $s$ can be pre-calculated exhaustively in $\Theta\left(\frac{1}{1-p_{skip}}\right)$ values.*

PROOF. Recognize that the expected value we desire is the number of trials we need to obtain a success, and so $1/p_{\text{skip}} = \mathbb{E}[s] = \sum_{i=1}^{\infty} p_{\text{skip}}^{i-1}(1 - p_{\text{skip}})^2$. This corresponds exactly to the Geometric distribution for trials till success, also known as the Pascal distribution [1], from which we can stably sample the value $s$, given $u \sim \mathcal{U}(0, 1)$:

$$s_u = \left\lfloor \frac{\log 1p(-u)}{\log(p_{\text{skip}})} \right\rfloor \quad (4)$$

However, unless $p_{\text{skip}}$ is very large, we will fail to amortize the cost of using floating point arithmetic. We remediate this problem by exploiting the floor operation $\lfloor \cdot \rfloor$ used in the sample calculation to perform a faster version of Walker's alias method [63, 64]. The standard alias method creates a table that is indexed using floating point arithmetic and is still known to be slow when many values or low latency are needed [30]. Instead, we can create a larger table $S$ of $|S|$ entries, where $S[i] \leftarrow s_{\frac{i+0.5}{|S|}}$. Necessarily, the last index of this table will not account for the exponentially decreasing chance of ever larger sample values — a non-issue in our application as it means we sample slightly more bytes than technically expected. With some careful algebra, we can set $|S|$ to guarantee extremely fast sampling and small errors.

First, we wish to find the smallest difference such that the value of the $s$ increases by one, and so solving the equation $\frac{\log\left(1-\frac{x+1}{|S|}\right)}{\log(p_{\text{skip}})} - \frac{\log\left(1-\frac{x}{|S|}\right)}{\log(p_{\text{skip}})} = 1$ yields

$$|S| = \frac{x \cdot p_{\text{skip}} - x - 1}{p_{\text{skip}} - 1}. \quad (5)$$

However, $x$ is not known, and so we take the CDF of the geometric distribution $1 - p_{\text{skip}}^{\lfloor x \rfloor}$ and replace the floor operation with $+\frac{1}{2}$ as an upper-bound that enables us to solve $1 - p_{\text{skip}}^{x+\frac{1}{2}} = c$, for any arbitrary $0 < c < 1$ as the desired fraction of the Geometric distribution we have covered. Solving this equation, we get:

$$x = \frac{2 \log(1 - c) - \log\left(p_{\text{skip}}\right)}{2 \log\left(p_{\text{skip}}\right)} \quad (6)$$

Substituting Equation 6 into Equation 5 we get

$$\frac{\frac{p_{\text{skip}}\left(2\log(1-c)-\log(p_{\text{skip}})\right)}{2\log(p_{\text{skip}})} - \frac{2\log(1-c)-\log(p_{\text{skip}})}{2\log(p_{\text{skip}})} - 1}{p_{\text{skip}} - 1}$$

which simplifies down to Equation 7

$$|S| = \frac{\log(1 - c)}{\log\left(p_{\text{skip}}\right)} + \frac{1}{1 - p_{\text{skip}}} - \frac{1}{2} \quad (7)$$

This guarantees that the table size $|S|$ will be sufficiently large such that every integer output from $[1, \text{CDF}^{-1}(c)]$ will be represented. However, this alone does not guarantee fast sampling due to the cost of generating random integers in the range of $[0, |S|)$. We instead round the value of Equation 7 up to the next power of two. via Equation 8. Doing so means we can cheaply sample the next 32 bit integer and shift right by $32 - \log_2(|S|)$ (special assembly functions exist to compute this in 1 cycle) bits to get the next value of $s \leftarrow S[\cdot]$. Thus, we can represent 99.9% of the Geometric distribution range for $p_{\text{skip}} \leq 99.9\%$ (i.e., all practical values in our experiments) within a cache of just 32 KB of cache (easily fits into L1 cache).

$$|S| = 2^{\left\lceil \frac{\log\left(\frac{\log(1-c)}{\log(p_{\text{skip}})} - \frac{1}{2} + \frac{1}{1-p_{\text{skip}}}\right)}{\log(2)} \right\rceil} \quad (8)$$

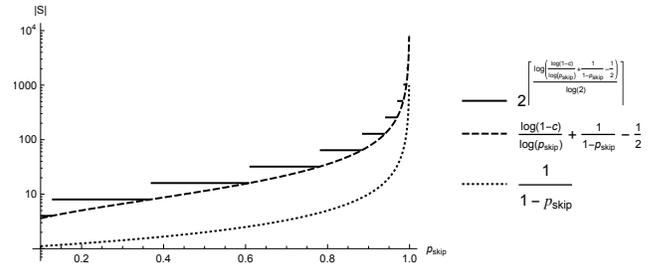

Figure 6: Our cache can be $\leq 8192$ items to accurately represent 99.9% of the Geometric distribution's range and handle a $p_{\text{skip}} \leq 99.9\%$, which is always the case in all of our experiments.

Note as well that the $\frac{1}{1-p_{skip}}$ term grows faster than $\frac{1}{\log p_{skip}}$, and so the value of $|S|$ grows at the rate of $\Theta\left(\frac{1}{1-p_{skip}}\right)$. □

Theorem 3's result, when we round the pre-calculate value cache size $|S|$ up to the next power of two, means we can cheaply sample the next 32-bit integer and shift right by $32 - \log_2(|S|)$ (special assembly functions exist to compute this in 1 cycle) bits to get the next value of $s \leftarrow S[\cdot]$. Doing so, we can represent 99.9% of the Geometric distribution range for $p_{\text{skip}} \leq 99.9\%$ (i.e., all practical values in our experiments) within a cache of just 32 KB of cache (easily fits into L1 cache). The 0.1% not represented is not a concern, as it is the tail of the distribution going toward $\infty$, and capping the 0.1% can only cause us to over-sample a (slightly) larger value of $N$.

### 4.3 Simple Write/Evict Cache

Our write/evict cache in Algorithm 1 is a simple data structure where we wish to accumulate the frequent updates into larger increments before reaching out to the main data structures. One may intuitively think that a Least Recently Used (LRU) or other kind of data structure would be appropriate for this task, but the overhead of these data structures is too great given the low overhead we can tolerate. Instead, a simple strategy of indexing into a small table and, on collision, evicting and replacing the previous value is used. By making this table the nearest power of two we can index into it using the rolling hash function quickly (avoiding a long latency modulo operation). One may hypothesize that such a naïve strategy would result in no benefit as the majority of items that occur less than once would overwhelm the cache, but we prove this strategy results in a log-linear reduction in main memory accesses (i.e., to the large hash-table or the SpaceSaving data structure).

THEOREM 4. *A cache of size $|C|$ reduces the number of writes to main memory by a factor of $\Theta(|C| \log |C|)$.*

PROOF. Let $f(\cdot)$ represent any PMF function. The likelihood of a collision in the cache, assuming uniformly distributed hashing, can be found $(1 - f(\cdot))/|C|$ by the pigeonhole principle. This creates the modified probability of the $x$'th item being evicted from the cache due to a collision $p_{\text{evict-}x} = \frac{f(x)}{f(x)+(1-f(x))/|C|}$. Given our Zipf assumption of $f(\cdot)$, we substitute the Zipf's PMF to obtain $\frac{1}{xH_\Sigma \left( \frac{1-\frac{1}{xH_\Sigma}}{C} + \frac{1}{xH_\Sigma} \right)}$, which simplifies to:

$$\frac{|C|}{|C| + xH_\Sigma - 1} \quad (9)$$

The average number of collisions for the $x$'th item is then the mean of the Geometric distribution, with probability Equation 9, giving $\mu_x = \frac{1}{1 - \frac{|C|}{|C|+xH_\Sigma-1}} - 1$. The average number of collisions saved can then be approximated as the sum of the top $|C|$ most probable items, $\sum_{x=1}^{|C|} \mu_x = \frac{|C| \left( H_{|C|-\frac{1}{H_\Sigma}} - H_{-\frac{1}{H_\Sigma}} \right)}{H_\Sigma}$. Taking the limit

$$\lim_{|C| \to \infty} \frac{\frac{|C| \left( H_{|C|-\frac{1}{H_\Sigma}} - H_{-\frac{1}{H_\Sigma}} \right)}{H_\Sigma}}{|C| \log(|C|)} = \frac{1}{H_\Sigma}$$

, which is a constant, thus concluding the proof. □

*Other Implementation Notes.* All tables are rounded up to the next power of two so that we can replace the modulo in a standard indexing operation (i.e., `i % length`) with bit-twiddling operations (i.e., `i >> bits`).

The first pass is parallelized using atomic updates. While parallel SpaceSaving varianets exist [5] they are non-trivial to implement and have non-trivial overhead. We instead create $P$ independently updated SpaceSaving data structures for $P$ processors and merge the final results (i.e., the "eps"' and "freq" counts in the verbiage of [31]) to produce a final rank list to select from.

## 5 Results

We now review the results of our method. First, we will establish that Zipf-Gramming has much higher computational throughput, especially compared to standard tools like Scikit-Learn. Second, we will establish that Zipf-Gramming results in similar accuracy on a variety of different file types of moderately sized corpora. Third, we will establish that Zipf-Gramming empirically captures the majority of true top-k features on a moderately sized corpus. Finally, only Zipf-Gramming can run on our production scale experiments, where we document significant gains in AUC and up to 76% absolute improvement in identifying challenging malware in our production environment.

Our results will be reported in AUC as our model is used at multiple different TPR vs. FPR thresholds, and we advise each downstream deployment to select a threshold applicable to their specific needs. For example, generic malware detection tasks will prefer a low FPR to avoid user issues. However, forensic investigations and other government tasks[3] often need maximal TPR, even when large FPR occur. Similarly, this is used as a component in larger systems as a faster pre-filter. This study is concerned only with the quality of this component, which has wide versatility in deployment.

### 5.1 Computational Throughput

In our first experiments, we examine the computational throughput of Zipf-Gramming and comparable baselines. In fairness to all prior approaches, we modify them as necessary to maximize their similarity to Zipf-Gramming only if the modification improves the MB/s of the prior approach. For example, the Kilo-Gramming implementation counts an n-gram only *once per file*, which causes a ≈ 10× slowdown as it must allocate a large dictionary, often larger in used memory than the file being processed due to overhead.

Our primary results are shown in Figure 7, where we see that computing the top 100,000 $n$grams for our Zipf-Gramming algorithm is 492× faster than Scikit-Learn when Scikit-Learn is given the benefit of using just 3-grams, as it could not successfully complete a run at 8-grams. These results were obtained running on a M3 Macbook Pro with 96 GB of RAM. Our method is the only one that can process larger corpora and continues to increase in efficiency as we do so. At $4.6 \cdot 10^{12}$ bytes of input, Zipf-Gramming reaches 2,762 MB/s, which would be 35.3× faster than Kilo-Grams if we could wait long enough for it to finish running. Scikit-Learn CountVectorizer[4] was tested using a 2 GB file for 6-grams. Using IO stats we observed bursts up to 63 KB/s, with multiple gaps of 0 KB/s, but the program crashed after several hours of computation.

It is worth noting that in the Kilo and Zipf-gramming algorithms, two passes over the dataset are made, and we are measuring the MB/s over the entire runtime of the program — not just the IO throughput per step. This means our Zipf-Gramming is reaching 5,524 MB/s IO throughput, and the maximum measured IO throughput for this machine when reading data to /DEV/NULL is 5777.7 MB/s, indicating we are at 95.61% of maximal practical efficiency and 73.65% of the theoretical maximum of 7,500 MB/s over the PCIe 4.0 interface used by our drive.

---
[3]https://afresearchlab.com/technology/successstories/cross-domain-solutions-101/
[4]https://scikit-learn.org/stable/modules/generated/sklearn.feature_extraction.text.CountVectorizer.html

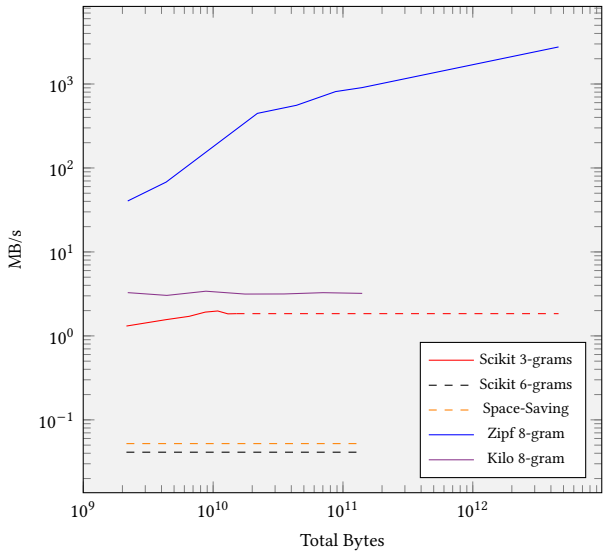

Figure 7: Our Zipf-Gramming is orders of magnitude faster (y-axis) than alternatives, and performance improves as dataset size (x-axis) increases. Dashed lines are not run to completion but measured with IO profiling tools due to exceeding a 1-hour timeout.

This also makes our approach particularly efficacious for any researcher developing on a local machine. Most NVMe connectors receive only four PCIe lanes, which would have a maximum theoretical throughput of 875 MB/s, and consumer SSDs connected via a SATA have a maximum theoretical throughput of 750 MB/s[5]. Our method handily surpasses what most individuals' workstations will be able to achieve. While developing a single-pass algorithm that can obtain higher theoretical runtime is desirable, only two-pass algorithms have thus far shown better aggregate IO throughput.

## 5.2 Equitable Quality

For our next experiment, we took several file types captured via VirusTotal, and used the ClarAvy tool [20] to determine benign and malicious labels. Not all file types are seen frequently or enough to build a production-quality model. This experiment is done purely to demonstrate that Zipf-Gramming produces models with similar final accuracy and that the results improve with the size of the dataset as the Zipf assumption holds. The five datasets and their AUC scores are shown in Table 2 are subsets of Ember3 before completion [21]. Each model was trained using Elastic-Net penalized logistic regression [66, 39, 44, 28, 29]. We see that Zipf-Gramming produces comparable feature sets but far faster, a requirement for timely updates to the production model.

---

[5]These are usually reported in Giga**bits** per second, instead of Giga**bytes** or Mega**bytes** per second, we have converted all to Megabytes per second to avoid confusion.

Table 2: Five datasets ordered by total corpus size (i.e., number of n-grams equals the number of bytes). Deterministic calculation of top-k n-grams is nearly equal to Zipf-Gramming performance but computationally infeasible for regular model updating at $\approx 100\times$ slower. Zipf-Gramming's resulting models improve as the dataset size increases.

| Dataset | Win32 | Win64 | ELF | .Net | PDF |
|---|---|---|---|---|---|
| Size | 2.54 TB | 1.85 TB | 328 GB | 189 GB | 45 GB |
| Kilo-Gramming | 95.73 | 96.48 | 99.29 | 96.48 | 82.98 |
| Zipf-Gramming | 95.44 | 96.37 | 99.26 | 94.63 | 81.59 |

## 5.3 Validating Top-k Accuracy

Having shown that our method is the fastest available for extracting the top-k most frequent $n$-grams, we next validate empirically that our theory is valid, and we obtain accurate estimates of the top-k items. This is distinct from showing accurate models, which is hypothetically possible with disjoint features. These experiments are carried out on real-world data, using a subset of Dot-Net benign and malicious programs totaling 188.93 GB in size as the maximum amount of data we could reasonably calculate the true top-k items. We perform this for 6-grams and 8-grams.

First, we look at the percentage of 6/8-grams missing from the top-k, and the relative error in the estimated count itself. For computational efficacy, this is measured by calculating the results for the top $k = 100,000$, and then using the ranked count order to evaluate all lower values of $k$. The results are shown below in Figure 8, where we can see that the percentage error is generally $\leq 2\%$.

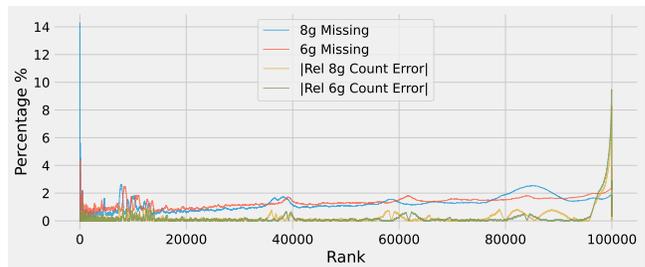

Figure 8: The percentage of missing top-k grams is small throughout the range of ranks (x-axis). The percent error in the estimated and true rank of an item in the $k$'th position is extremely small, usually $\leq 10^{-3}\%$, and increases to a still inconsequential error rate at the very last $k$ items.

The red and blue lines are the rates of missing grams from the top-$k$, which peaks at the lowest ranks of $k \leq 100$ due to errors in order and placement. This is primarily an artifact of using real-world data that are not truly I.I.D. distributed in the occurrence of n-grams, but the low error rate validates the theoretical assumption of Zipfian distributed features.

A key to the acceptability of the approximation by sub-sampling is that even if we make an error, the errors are inconsequential. For example, the $k + 1$'th most frequent item being selected instead of the $k$'th is unlikely to have an impact on model accuracy due to similar occurrence rates. The yellow and green lines of Figure 8 show that the estimated count error of the 6/8-gram is very nearly 0 in almost all cases, with the error rate spiking only at the very bottom $k$ ranks but at a range $\leq 10\%$ indicating that these errors are inconsequential as we are selecting features with highly similar occurrence rates. If we had selected a problematic feature instead, such as one that was an order of magnitude further down the rank list, we would expect to see higher relative error rates.

We can further validate this by looking at the error in the rank value itself, i.e., how far away from the true rank of the $k$'th item is from the Zipf sub-sampled observed rank. This is plotted in Figure 9 with the mean and standard deviation, where we see, as expected, that the average rank error is zero, but the standard deviation of the order increases as we descend to less frequent items. The overall distribution in the figure is measured as a moving window over 1,000 ranks.

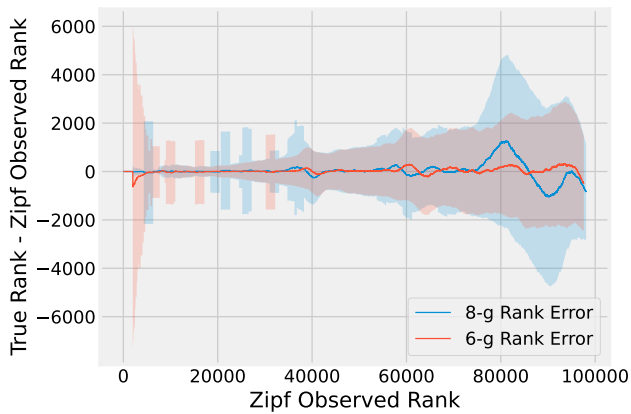

Figure 9: The difference in true and observed rank has a mean error of zero, but the variance in rank-order errors increases as we descend toward the tail of the distribution. A small number of outliers exist at the top ranks, causing asperity to the variance.

Note in the high ranks, there are bursts of high variance caused by singleton outliers that are mis-ranked by large values, which occurs due to the non-I.I.D. distribution that occurs in real data. For example, the 6-gram 0x909090909090 is very common as the "No Operation" instruction (0x9) is repeated numerous times in a file for a variety of benign and malicious applications. By the nature of it's use, such occurrences are sequential in occurrence and thus clumped together instead of spread randomly throughout files. In such circumstances, our Zipfian approach is guaranteed to undercount the true occurrence rate of the feature. However, this does not happen with sufficient frequency to meaningfully change the results, and the mean error in rank is 11 and 16 for 6-grams and 8-grams, respectively.

## 5.4 Deployment Results

We now turn to the results of our method in deployment. Our previous model used byte 8-grams via the Kilo-Gramming algorithm for $k$ = 1 million, and used Elastic-Net penalized logistic regression to build the model and down-select the number of relevant features to a more compact and deployable subset. This model was trained on approximately 2.5 million files from a proprietary corpus. Building this model required three to four days of compute time, and increasing the training set size had proven to be a cumbersome roadblock toward improvement. In internal testing, we observed that we could not simply replace all old data with new data either, as identifying old malware is still operationally useful.

Our new model was trained on an additional 2.6 million files, giving a total of 5.1 million files, and is 6.6 TB in size. Using the Zipf-Gramming approach, we extracted the top $k$ = 10 million 6-grams, followed by Elastic-Net penalized logistic regression, which selected only 33 thousand features for the final model. *Despite our attempts to use this with the original Kilo-Gramming code, the estimated runtime to completion was 1 month, which was not viable fiscally or in terms of producing regularly updated models*. Developing the Zipf-gramming approach was what allowed us to produce new models of improved quality and with a runtime of just *four hours* on this corpus, is well within our ability to perform regular updates.

Multiple tests are used to evaluate our approach, each labeled using the aggregate results from Anti-Virus products in VirusTotal [62], using ClarAVy [20] to aggregate the reported output of each AV system to determine the benign/malicious label. These labels are high quality but can only be obtained months after the fact due to relying on AntiVirus outputs stabilizing. *We rely on them to retrospectively confirm our model's efficacy in a quantified way*.

First is a "Standard Test" set composed of data in a 3-month time span after the last observed data point in the training set. This is done to avoid test-set leakage due to correlations in time in the training data [37]. Second, we collect a "GAP" dataset that contains 132,700 samples collected 9 months after deployment, relevant to the Standard Test set, imposing a greater strain on the model. Third, we note that AV outputs change over time as they are updated, and we used a "Challenge" set that includes only malware sampled in an adversarial manner to be maximally difficult. Each Challenge file occurred after the original training data and was initially labeled as benign by all available AV products in VirusTotal when first seen, but was later labeled malicious by a plurality of AV products 30 days later. This indicates a sample of malware that is uniquely challenging because it was sufficiently different to be undetected when first deployed and only captured after the fact when AV systems made new updates. In this challenge set, there are 2533 Win32, 822 Win64, and 708 .Net files.

Our results in Table 3 show that increasing the datasize massively improves performance, especially on challenging files that represent zero-day malware. This training size was not viable without improved feature extraction efficiency, and allows us to continue deploying a small (2MB) model that can run efficiently on many platforms via CPU only, as GPUs are not available or even compatible with many needed systems.

We note that our original intention was to include the publicly released Sophos dataset, which in total has 19,389,877 files, roughly split between benign and malicious. This dataset far exceeds the feasibility of the original Kilo-Gramm algorithm and code. Our Zipf-Gramming algorithm processed the entire corpus in 24 hours and 37 minutes. However, we found that including the Sophos data decreased model performance on our test datasets, which we believe most represent the range of malware we care about. To illustrate this, we trained a 6-gram model on the top $k$=10 million, with the results in the table Table 4. Despite being much larger in size, we see uniformly worse performance, especially in our Challenge tests.

Table 3: Our new model performs better in our standard train/test split (first rows), on 9 months of proceeding production data (middle rows), and on a set of challenge malware that was missed by all Anti-Virus products (bottom rows).

|     | Standard Test | | Gap | | | Challenge | | |
| --- | --- | --- | --- | --- | --- | --- | --- | --- |
|     | F1 | AUC | Win32 | Win64 | .Net | Win32 | Win64 | .Net |
| New | 89.70 | 96.00 | 96.15 | 94.58 | 94.59 | 49.03 | 36.04 | 79.94 |
| Old | 73.43 | 91.35 | 92.02 | 64.43 | 97.60 | 32.45 | 0.81 | 0.14 |

Table 4: The Sophos data, despite larger in size, performed unexpectedly worse across all metrics compared to our curated internal dataset.

|     | Standard Test | | Gap | Challenge | | |
| --- | --- | --- | --- | --- | --- | --- |
|     | F1 | AUC | AUC | Win32 | Win64 | .Net |
| New Model | 89.70 | 96.00 | 95.19 | 49.03 | 36.04 | 79.94 |
| Sophos 6-g | 80.67 | 85.40 | 84.37 | 1.55 | 4.26 | 3.52 |

## 6 Conclusion

Byte n-grams are the only feature allowing us to satisfy a runtime/footprint critical use case for malware detection. By subsampling the data under the Zipf distribution assumption, we can carefully design a system thousands of times faster than Scikit-Learn, which enables us to produce a production-grade model in one day rather than a month as before. This improved our production model's identification of evasive challenging malware by up to 70% absolute points and was validated theoretically and empirically.

### GenAI Usage Disclosure section

No generative AI of any kind was used at any point for this work.